\documentclass[aps,prl,showpacs,floatfix,twocolumn,byrevtex,superscriptaddress]{revtex4-1}

\usepackage{amsmath}
\usepackage{amssymb}
\usepackage{amstext}
\usepackage{amsopn}
\usepackage{amsfonts}
\usepackage{amsxtra}
\usepackage[english]{babel}
\usepackage{graphicx}
\usepackage{float}
\usepackage{bm}
\usepackage{multirow}
\usepackage{dcolumn}
\usepackage{hyperref}
\usepackage{color}

\newcommand{\dr}{$\Delta R(t)/R$}
\newcommand{\wt}{WTe$_{2}$}
\newcommand{\ep}{$e$-ph}
\newcommand{\ee}{$e$-$e$}
\newcommand{\eh}{$e$-$h$}

\begin{document}

\title{Ultrafast Carrier Dynamics in the Large Magnetoresistance Material WTe$_{2}$}
\author{Y. M. Dai}
\author{J. Bowlan}
\affiliation{Center for Integrated Nanotechnologies, Los Alamos National Laboratory, Los Alamos, New Mexico 87545, USA}
\author{H. Li}
\author{H. Miao}
\author{S. F. Wu}
\author{W. D. Kong}
\author{Y. G. Shi}
\affiliation{Beijing National Laboratory for Condensed Matter Physics, Institute of Physics, Chinese Academy of Sciences, P.O. Box 603, Beijing 100190, China}
\author{S. A. Trugman}
\author{J.-X. Zhu}
\affiliation{Center for Integrated Nanotechnologies, Los Alamos National Laboratory, Los Alamos, New Mexico 87545, USA}
\affiliation{Theoretical Division, Los Alamos National Laboratory, Los Alamos, NM 87545, USA}
\author{H. Ding}
\affiliation{Beijing National Laboratory for Condensed Matter Physics, Institute of Physics, Chinese Academy of Sciences, P.O. Box 603, Beijing 100190, China}
\affiliation{Collaborative Innovation Center of Quantum Matter, Beijing, China}
\author{A. J. Taylor}
\author{D. A. Yarotski}
\author{R. P. Prasankumar}
\email[]{rpprasan@lanl.gov}
\affiliation{Center for Integrated Nanotechnologies, Los Alamos National Laboratory, Los Alamos, New Mexico 87545, USA}

\date{\today}

%
%

\begin{abstract}
Ultrafast optical pump-probe spectroscopy is used to track carrier dynamics in the large magnetoresistance material WTe$_{2}$. Our experiments reveal a fast relaxation process occurring on a sub-picosecond time scale that is caused by electron-phonon thermalization, allowing us to extract the electron-phonon coupling constant. An additional slower relaxation process, occurring on a time scale of $\sim$5-15 picoseconds, is attributed to phonon-assisted electron-hole recombination. As the temperature decreases from 300~K, the timescale governing this process increases due to the reduction of the phonon population. However, below $\sim$50~K, an unusual decrease of the recombination time sets in, most likely due to a change in the electronic structure that has been linked to the large magnetoresistance observed in this material.
\end{abstract}



\maketitle

%
%

The recent discovery of large positive magnetoresistance (MR) in a number of nonmagnetic materials such as WTe$_{2}$~\cite{Ali2014}, Cd$_{3}$As$_{2}$~\cite{Liang2015} and NbSb$_{2}$~\cite{Wang2014} has aroused tremendous interest, due not only to its potential application in devices such as magnetic sensors and hard drives, but also to the enigmatic nature of this effect. The observed MR in Cd$_{3}$As$_{2}$ has been attributed to the recovery of backscattering that is strongly suppressed in zero magnetic field~\cite{Liang2015}. In WTe$_{2}$, the large non-saturating MR is believed to arise from perfect electron-hole ($e$-$h$) compensation~\cite{Ali2014}, similar to bismuth (Bi) and graphite~\cite{Alers1953,Yang1999,Du2005}. This is supported by angle-resolved photoemission spectroscopy (ARPES)~\cite{Pletikosic2014} and quantum oscillation~\cite{Zhu2015} experiments, which have found hole and electron pockets with the same size in \wt\ at low temperatures. Further support for the $e$-$h$ compensation scenario came from the application of pressure, which increases the difference between the sizes of the hole and electron pockets, dramatically suppressing the MR~\cite{Cai2015,Kang2015,Pan2015}. However, a recent high resolution ARPES study revealed a more complicated Fermi surface (FS) with nine pockets~\cite{Jiang2015}. In addition, circular dichroism was observed in the photoemission spectra, signaling strong spin-orbit coupling, which may also play an important role in \wt. Finally, a detailed study of the Shubnikov-de-Haas (SdH) effect, in combination with density functional theory (DFT) calculations, indicated that perfect $e$-$h$ compensation breaks down under an external magnetic field in \wt~\cite{Rhodes2015}, challenging the $e$-$h$ compensation scenario.

More insight into the physics of WTe$_{2}$ can be obtained using ultrafast optical spectroscopy, which tracks the relaxation of photoexcited carriers in the time domain as they return to equilibrium. Carrier relaxation depends sensitively on the band structure and scattering mechanisms in a solid~\cite{Torchinsky2010,Demsar2003,Chia2010,Qi2013}, and therefore an understanding of the ultrafast carrier dynamics in \wt\ may shed new light on the mechanism of the anomalous MR. This also directly resolves the timescales that ultimately limit potential applications of \wt\ in electronic devices, e.g., high speed solid-state drives. However, ultrafast optical studies have not yet been performed on this material.

Here, we present a detailed ultrafast transient reflectivity (\dr) study on \wt\ while varying the temperature ($T$) and photoexcitation fluence ($F$). Two distinct relaxation processes contribute to the recovery of the photoinduced change in reflectivity. We find that electron-phonon (\ep) thermalization is responsible for the fast decay that occurs on a sub-picosecond (ps) time scale, which allows us to extract the \ep\ coupling constant, an important material parameter, using a two-temperature model. Phonon-assisted $e$-$h$ recombination between the electron and hole pockets then takes place within $\sim$5-15~ps, giving rise to the slow decay in the \dr\ signal. This process gradually slows down upon cooling due to the decreasing phonon population, but then becomes faster again below $\sim$50~K, which is likely caused by an electronic structure change that has been linked to the large MR in \wt.

%
%

High-quality \wt\ single crystals were grown by solid-state reactions using Te as the flux. The starting materials W (99.9\%, General Research Institute For Nonferrous Metals, Beijing, China) and excessive amounts of Te (99.999\%, Alfa Aesar, U.S., lot. 10362) were mixed and placed in an alumina ampoule, then sealed in a quartz tube. The operation was conducted in a glove box filled with high-purity argon gas. The temperature was increased to 1000$^{\circ}$C over 10~h, after which the sample was heated in a furnace at 1000$^{\circ}$C for 5~h. After reaction, the system was cooled down to 800$^{\circ}$C at a rate of 1$^{\circ}$C/h, then to 700$^{\circ}$C in 20~h. The quartz tube was then inverted and quickly spun in a centrifuge to remove the Te flux.

The transient reflectivity measurements were carried out using a Ti:sapphire femtosecond (fs) laser oscillator producing pulses with a center wavelength of 830~nm (1.5~eV), a duration of 40~fs and a repetition rate of 80~MHz. The spot sizes of the pump and probe beams at the sample were $\sim$65 and 32~$\mu$m in diameter, respectively. For $F$ = 3.5~$\mu$J/cm$^2$, using the optical properties at 830~nm given in ref. \cite{Homes2015}, the photoexcited carrier density in \wt\ is estimated to be $\sim$3.24$\times 10^{18}$~cm$^{-3}$. Data was collected on a freshly cleaved surface from 5 to 300~K, with the pump and probe beams cross-polarized.

%
%

\begin{figure}[tb]
\includegraphics[width=0.8\columnwidth]{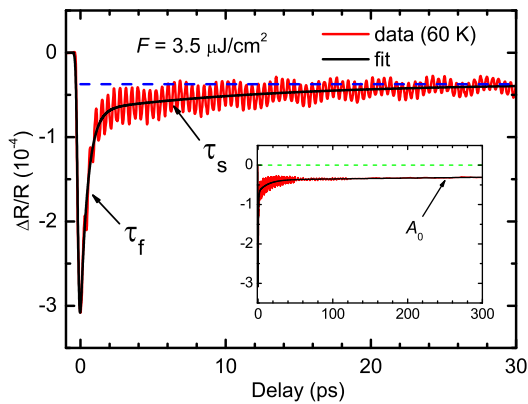}
\caption{ (color online) The transient reflectivity \dr\ (red line) as a function of time delay in WTe$_{2}$ at $T$ = 60~K and $F$ = 3.5~$\mu$J/cm$^{2}$ . The black solid line shows the fit to the measured data, without including the oscillations. The dashed blue line shows the value of the \dr~signal at long time delays, $A_{0}$. The inset shows the \dr\ signal (red line) and the fitting result (black line) up to 300~ps.}
\label{Fig1}
\end{figure}

Figure~\ref{Fig1} shows the \dr\ trace for WTe$_{2}$ at $T$ = 60~K and $F$ = 3.5~$\mu$J/cm$^{2}$.
%
Photoexcitation results in a sharp drop in \dr\ due to the change in temperature of the electronic system after electron-electron ($e$-$e$) thermalization, followed by two distinct relaxation processes. A fast recovery of the photoinduced reflectivity change ($\tau_{f}$) occurs within 1~ps, after which a slow decay ($\tau_{s}$) with a weaker amplitude is observed. At longer time scales, the \dr\ signal is dominated by a flat offset ($A_{0}$), as shown in the inset of Fig.~\ref{Fig1}. In addition to these two decay processes, two oscillations with different frequencies are clearly observed in the \dr\ trace. These oscillations originate from the generation of  coherent lattice vibrations by the femtosecond pump pulse~\cite{Hase2005,Qi2010}. The frequencies of the two oscillations are 0.25 and 2.4~THz, respectively, agreeing well with the energies of two A$_{1}$ optical phonon modes in \wt~\cite{Kong2015,Jiang2015a}.

\begin{figure}[tb]
\includegraphics[width=\columnwidth]{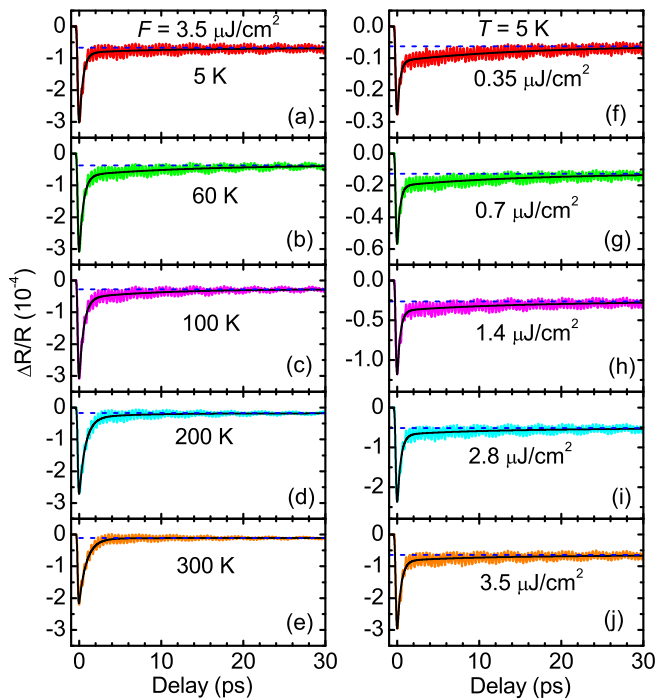}
\caption{ (color online) (a)--(e) \dr\ in \wt\ at several selected temperatures between 5 and 300~K. The pump fluence is $F$ = 3.5~$\mu$J/cm$^{2}$. (f)--(j) \dr\ in \wt\ measured with different pump fluences at 5~K. The black solid line through the data in each panel denotes the curve fit to the non-oscillatory response.}
\label{Fig2}
\end{figure}

In this Letter, we will focus on the non-oscillatory response, \emph{i.e.} the two relaxation processes, which provides pivotal information about the ultrafast dynamics of photoexcited carriers in \wt. To gain insight into the origin of the two relaxation processes, a careful examination of the $T$ and $F$ dependence of $\tau_{f}$ and $\tau_{s}$ is indispensable. Figures ~\ref{Fig2}(a)--\ref{Fig2}(e) depict \dr\ at selected temperatures between 5 and 300~K. This reveals that $\tau_{f}$ becomes faster as the temperature is reduced; $\tau_{s}$ also exhibits noticeable $T$ dependence, as can be seen by comparing the \dr\ curve with the blue dashed line ($A_{0}$) in each panel. Figures~\ref{Fig2}(f)--\ref{Fig2}(j) show \dr\ measured with different fluences at 5~K. While $\tau_{f}$ is approximately $F$ independent, $\tau_{s}$ becomes faster with increasing pump fluence.

\begin{figure}[tb]
\includegraphics[width=0.8\columnwidth]{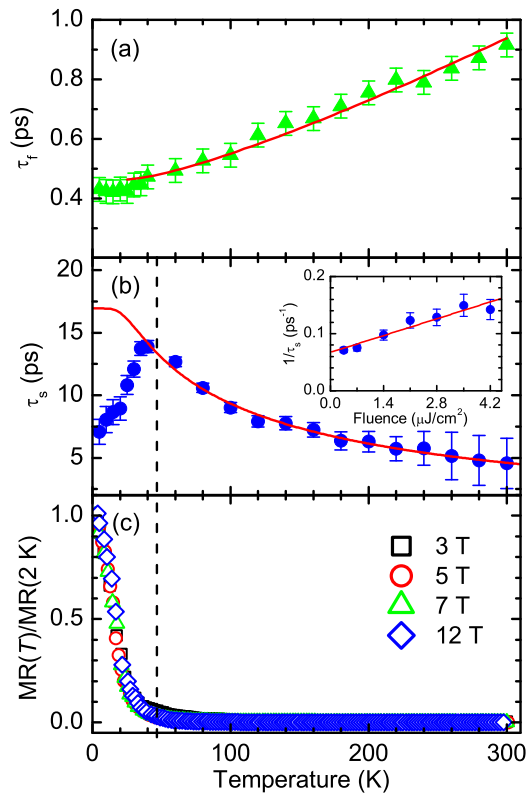}
\caption{ (color online) (a) and (b) show $\tau_{f}$ and $\tau_{s}$, respectively, as a function of temperature for $F = 3.5$~$\mu$J/cm$^{2}$. The red solid curve denotes (a) the TTM fit and (b) the calculated $T$ dependence of the phonon-assisted $e$-$h$ recombination time, with a fixed density of states in the electron and hole Fermi pockets. The inset of (b) displays the decay rate $1/\tau_{s}$ as a function of the pump fluence, while the red solid line represents a linear fit to the data. (c) Temperature dependence of the MR, normalized by the values at 2~K.}
\label{Fig3}
\end{figure}

In order to quantitatively study the $T$ and $F$ dependence of $\tau_{f}$ and $\tau_{s}$, we extract their time constants by fitting the measured \dr\ signal to two exponential decays convoluted with a Gaussian function (representing the autocorrelation of the pump and probe pulses) at all measured temperatures and fluences. We also included the long time offset $A_{0}$, which is usually related to heat diffusion out of the excitation volume on a timescale of several nanoseconds~\cite{Luo2012,Cheng2014}, much longer than the measured timescale (300~ps). Therefore, we can reasonably take it as a constant. Figures~\ref{Fig3}(a) and \ref{Fig3}(b) display the detailed $T$ dependence of $\tau_{f}$ and $\tau_{s}$, respectively. The decay rate for the slow component, $1/\tau_{s}$, is plotted as a function of $F$ in the inset of Fig.~\ref{Fig3}(b).

%

Having examined the $T$ and $F$ dependence of the two decay processes, we next trace their origins. In most materials, after femtosecond photoexcitation and $e$-$e$ thermalization, which give rise to the initial sharp change in the \dr~signal, the hot carriers transfer their excess energy to the lattice on a sub-ps time scale through \ep~interactions~\cite{Allen1987,Groeneveld1995,Prasankumar2011,Cheng2014}. This makes it reasonable to attribute $\tau_{f}$ in \wt\ to the cooling of the electronic system via \ep~ thermalization. Further support comes from the $T$ dependence of $\tau_{f}$ [Figure~\ref{Fig3}(a)], which is quite similar to the $T$-dependent \ep\ thermalization time observed in other materials~\cite{Groeneveld1995,Hase2005,Cheng2014}.
%

Electron-phonon thermalization is generally described by the two-temperature model (TTM)~\cite{Allen1987,Demsar2003,Cheng2014}. Within this model, the \ep\ thermalization time $\tau_{e-ph}$ is given by~\cite{Groeneveld1995,Cheng2014}
\begin{equation}
\tau_{e-ph} = \frac{\gamma(T_{e}^{2} - T_{l}^{2})}{2 H(T_{e},T_{l})},
\label{TTM}
\end{equation}
where $\gamma$ is the electron specific heat coefficient. $T_{e}$ corresponds to the electron temperature after \ee\ thermalization, which is initially higher than the lattice temperature $T_{l}$. $H(T_{e},T_{l})$ takes the form
\begin{equation}
H(T_{e},T_{l}) = f(T_e) - f(T_l),
\label{H}
\end{equation}
where
\begin{equation}
f(T) = 4g_{\infty} \frac{T^5}{\theta_{D}^4} \int_{0}^{\theta_{D}/T} \frac{x^4}{e^x - 1} dx,
\label{f}
\end{equation}
with $\theta_{D}$ and $g_{\infty}$ denoting the Debye temperature and the \ep~coupling constant, respectively. $T_{e}$ can be calculated using
\begin{equation}
T_e = \left( T_{l}^{2} + \frac{2U_{l}}{\gamma} \right)^{1/2},
\label{Te}
\end{equation}
where $U_l$ is the deposited laser energy density. It is well known that the TTM is not an appropriate model for \ep\ thermalization at low temperatures ($T \leq \theta_{D}/5$), where the $e$-$e$ and \ep~thermalization times become comparable~\cite{Groeneveld1995,Demsar2003,Hase2005}. However, since $\theta_{D}$ = 133.8~K for \wt~\cite{Callanan1992}, it is reasonable to use the TTM to describe \ep~thermalization above $\sim$30~K. The red solid curve in Fig.~\ref{Fig3}(a) shows that the TTM fits the $T$ dependence of $\tau_{f}$ (green solid triangles) quite well above 30~K, which supports our assignment of $\tau_{f}$ to \ep\ thermalization. Using the optical properties measured by Homes \emph{et al.}~\cite{Homes2015}, the deposited laser energy density is determined to be $U_l \simeq 0.78$~J/cm$^3$. Consequently, we can obtain the \ep\ coupling constant and the electronic specific heat coefficient from the TTM fit: $g_{\infty} \simeq$~6.79~$\times$~10$^{15}$~W~m$^{-3}$~K$^{-1}$ and $\gamma \simeq$~0.81~mJ~mol$^{-1}$~K$^{-2}$. The \ep\ coupling constant determined here will be important in understanding the mechanism of the recently discovered superconductivity in \wt\ under pressure~\cite{Kang2015,Pan2015}, underlining the utility of ultrafast optical spectroscopy for extracting fundamental parameters in complex materials.

\begin{figure}[tb]
\includegraphics[width=0.6\columnwidth]{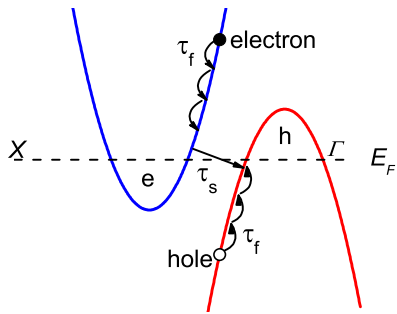}
\caption{ (color online) Schematic band structure of \wt\ near the Fermi level ($E_{F}$) along the $\Gamma$-$X$ direction at low temperatures. $\tau_{f}$ describes the \ep\ thermalization process and $\tau_{s}$ represents the phonon-assisted interband \eh\ recombination process.}
\label{Fig4}
\end{figure}

The origin of the slow decay ($\tau_{s}$) may be uncovered by examining the FS and electronic band structure of \wt. DFT calculations~\cite{Augustin2000,Ali2014} have shown that both the valence and conduction bands barely cross the Fermi level at different points in the Brillouin zone, producing a pair of electron and hole pockets with equal size on each side of the $\Gamma$ point in the Brillouin zone. This FS structure has been confirmed by ARPES~\cite{Pletikosic2014} and quantum oscillation~\cite{Cai2015,Zhu2015} experiments. A recent detailed ARPES measurement revealed a more complex FS~\cite{Jiang2015}. Nevertheless, pairs of electron and hole pockets represent the dominant feature of the FS structure. Based on the above theoretical and experimental studies, a schematic band structure in the $\Gamma$-$X$ direction for \wt\ at low temperatures is traced out in Fig.~\ref{Fig4}, which will help us to understand the origin of $\tau_{s}$.

%
After \ep\ thermalization ($\tau_{f}$) takes place (schematically illustrated in Fig.~\ref{Fig4}), there are still excess electrons (holes) in the conduction (valence) band. $\tau_{s}$ may therefore arise from $e$-$h$ recombination between the conduction and valence bands. This could occur radiatively, as in semiconductors, where an electron in the conduction band can recombine with a hole in the valence band by emitting a photon~\cite{Othonos1998}. However, this is unlikely to be the origin of $\tau_{s}$ in \wt, because (i) radiative recombination usually occurs on a much longer time scale (a few nanoseconds)~\cite{Othonos1998}; (ii) the emission of a photon cannot provide the momentum transfer needed for $e$-$h$ recombination in \wt. Alternatively, $e$-$h$ recombination could occur via a three-body Auger process~\cite{Vodopyanov1992,Onishi2015}, where an electron and a hole recombine and transfer the resulting energy to a third charge carrier. Nonetheless, (i) the $T$ dependence of $\tau_{s}$ [Fig.~\ref{Fig3}(b)] does not favor Auger recombination, since this process is generally insensitive to temperature; (ii) the linear dependence of the decay rate $1/\tau_{s}$ on $F$, as shown in the inset of Fig.~\ref{Fig3}(b), is a clear signature of two-particle dynamics~\cite{Segre2002,Torchinsky2010}.

Finally, we consider phonon-assisted $e$-$h$ recombination, where the electron and hole recombine with the assistance of \ep\ scattering between the electron and hole pockets, as illustrated by $\tau_{s}$ in Fig.~\ref{Fig4}. This has previously been observed and discussed in Bi~\cite{Lopez1968,Sheu2013}, a semimetal with a similar band structure to \wt. The $e$-$h$ recombination time in bismuth was found to be strongly $T$ dependent and varies from a few to tens of ps~\cite{Lopez1968,Sheu2013}, in good agreement with the features of $\tau_{s}$. Furthermore, we note that in \wt, the conduction and valence band extrema are in close proximity within the Brillouin zone~\cite{Ali2014,Pletikosic2014}, making interband \ep\ scattering more favourable. These facts indicate that $\tau_{s}$ in \wt\ may be due to phonon-assisted $e$-$h$ recombination between the electron and hole pockets.

Both theoretical and experimental studies in graphene and Bi have shown that the phonon-assisted $e$-$h$ recombination time strongly depends on temperature~\cite{Rana2009,Lopez1968}, because the phonon population is a strong function of temperature. Lowering the temperature reduces the efficiency of interband \ep\ scattering, leading to a longer $e$-$h$ recombination time. This is exactly what we observed in \wt\ at high temperatures (above $\sim$50~K) as shown in Fig.~\ref{Fig3}(b). In the presence of \ep\ coupling, the $T$ dependence of the phonon-assisted $e$-$h$ recombination time, $\tau_R$, can be quantitatively described using a model previously applied to phonon-assisted \eh\ recombination in Bi~\cite{Lopez1968}:
%
\begin{equation}
\frac{1}{\tau_R} =  A \frac{\frac{\hbar \omega}{2kT}}{\sinh^{2}(\frac{\hbar \omega}{2kT})} + \frac{1}{\tau_{0}},
\label{ehrt}
\end{equation}
where $\omega$ corresponds to the frequency of the phonon mode assisting \eh\ recombination, $\tau_{0}$ represents a $T$-independent recombination time which varies with sample purity, and $A$ is a parameter related to the density of states in the conduction and valence bands and the matrix element for interband \ep\ scattering. The red curve in Fig.~\ref{Fig3}(b) is the least-squares fit to the data using Eq.~(\ref{ehrt}). The excellent agreement between the experimental data and the fitting result above 50~K strongly supports the assignment of $\tau_{s}$ to phonon-assisted \eh\ recombination.

However, below 50~K, an unusual decrease in $\tau_{s}$ sets in and continues down to 5~K, strikingly deviating from the expected $T$ dependence of phonon-assisted \eh\ recombination [red curve in Fig.~\ref{Fig3}(b)]. Generally, a sharp change in the relaxation time constant often results from a phase transition~\cite{Demsar1999,Chia2010}, yet no phase transition has been reported at this temperature in \wt~\cite{Ali2014,Pletikosic2014,Kong2015,Zhu2015,Callanan1992,Kabashima1966}. On the other hand, there is a growing body of evidence indicating that a change in the electronic structure happens at $\sim$50 K. For example, a recent transport study revealed a sharp enhancement of the mass anisotropy below $\sim$50~K, which has been attributed to a temperature-induced change in the electronic structure~\cite{Thoutam2015}. In addition, the ARPES study in ref.~\cite{Pletikosic2014} mapped the Fermi surfaces of \wt\ at 20 and 100~K, revealing that the hole pocket expands dramatically at low temperatures. Finally, the observations of a sign change in the thermoelectric power at $\sim$60~K~\cite{Kabashima1966} and a sharp increase of the Hall coefficient below $\sim$50~K in \wt~\cite{Luo2015} may also point to an electronic structure change.

A comparison between our experimental results and the above observations therefore suggests that the decrease in $\tau_{s}$ below 50~K is most likely associated with a change in the electronic structure. This is consistent with the fact that phonon-assisted \eh\ recombination is sensitive to the density of states in the conduction and valence bands as well as the matrix element for interband \ep\ scattering~\cite{Lopez1968}, both of which are closely tied to the electronic structure of the material. In \wt, $\tau_{s}$ may be influenced by the electronic structure in the following ways: (i) expansion of the hole pocket can accelerate \eh\ recombination, causing $\tau_{s}$ to decrease at low temperatures, since there are more holes for the photoexcited electrons to recombine with~\cite{Rana2009}; (ii) an electronic structure change may also affect the matrix element for interband \ep\ scattering, thus modifying the phonon-assisted \eh\ recombination rate.

Finally, by comparing $\tau_{s}$ [Fig.~\ref{Fig3}(b)] with the $T$ dependence of the normalized MR [Fig.~\ref{Fig3}(c)] measured on our sample, we found that the MR turns on at the point where $\tau_{s}$ starts to fall, and increases sharply as the deviation of $\tau_{s}$ from the expected behavior grows. This strongly suggests that the large MR and the anomalous behavior of $\tau_{s}$ below $\sim$50~K are correlated. More importantly, this indicates that the electronic structure change, which gives rise to the anomalous behavior of $\tau_{s}$ at low temperatures, may also be responsible for the ``turn on'' effect of the large MR in \wt. A similar conclusion was reached by Thoutam \emph{et al.} in their recent transport study~\cite{Thoutam2015}.

%
%

In conclusion, we performed ultrafast transient reflectivity measurements on \wt\ as a function of pump fluence and temperature. Two relaxation processes were clearly observed. We have ascribed the fast decay process ($\tau_{f}$) to carrier cooling through \ep\ thermalization, allowing us to use a two-temperature model to extract the \ep\ coupling constant and the electronic specific heat coefficient. The slow decay process ($\tau_{s}$) is due to phonon-assisted $e$-$h$ recombination between the electron and hole bands. While $\tau_{s}$ follows the expected $T$ dependence quite well down to $\sim$50~K, a pronounced deviation of $\tau_{s}$ from the expected behavior appears at low temperatures. A comparison with other experimental results indicate that the anomalous $T$ dependence of $\tau_{s}$ below $\sim$50~K is due to an electronic structure change, which likely plays a key role in promoting the large MR in \wt.

%
%

We thank Pamela Bowlan, Xia Dai, Jiangping Hu, Yongkang Luo, Brian McFarland, Kamaraju Natarajan, Ivo Pletikosi\'{c}, Qiang Wang and Xianxin Wu for helpful discussions and especially Christopher C. Homes for sharing his unpublished optical data. This work was performed at the Center for Integrated Nanotechnologies, a U.S. Department of Energy, Office of Basic Energy Sciences user facility. Los Alamos National Laboratory is operated by Los Alamos National Security, LLC, for the National Nuclear Security administration of the U.S. Department of Energy under contract no. DE-AC52-06NA25396. Work at LANL was supported by the LANL LDRD program and by the UC Office of the President under the UC Lab Fees Research Program, Grant ID No. 237789. Work at IOP CAS was supported by the Strategic Priority Research Program (B) of the Chinese Academy of Sciences (Grant No. XDB07020100) and the National Natural Science Foundation of China (No. 11274367, 11474330).

%
%

\end{document}